\documentclass[10pt,journal,compsoc]{IEEEtran}

\usepackage{amsmath,amssymb}
\usepackage{booktabs}
\usepackage{array}
\usepackage{graphicx}
\usepackage{url}
\usepackage[hidelinks]{hyperref}
\usepackage{xcolor}
\usepackage{colortbl}
\usepackage{multirow}
\usepackage{balance}

\definecolor{v27g}{RGB}{223,239,223}
\definecolor{v26r}{RGB}{247,220,220}
\definecolor{splity}{RGB}{250,242,214}
\definecolor{hdr}{RGB}{31,78,121}

\newcommand{\gcell}{\cellcolor{v27g}}
\newcommand{\rcell}{\cellcolor{v26r}}

\begin{document}

\title{Quantifying Bitcoin Network Resilience Through Critical Scenario Discovery: A Dual-Layer Framework for Discovering Contentious Fork Conditions in Decentralized Consensus}

\author{Peter~Foytik, Sachin~Shetty, Ross~Gore, and Eranga~Bandara%
\thanks{P. Foytik, S. Shetty, R. Gore, and E. Bandara are with the Center for Secure and Intelligent Critical Systems (CSICS), Old Dominion University, Norfolk, VA, USA (e-mail: pfoytik@odu.edu; sshetty@odu.edu; rgore@odu.edu; cmedawer@odu.edu).}%
\thanks{Preprint. A condensed version of a working draft prepared for the University of Wyoming Bitcoin Research Institute Workshop, July 2026.}}

\markboth{Preprint - July 2026}{Foytik \MakeLowercase{\textit{et al.}}: Quantifying Bitcoin Network Resilience Through Critical Scenario Discovery}

\IEEEtitleabstractindextext{%
\begin{abstract}
Bitcoin's consensus depends not only on protocol rules but on the emergent behavior of a heterogeneous network of independently configured nodes with divergent economic stakes. Whether a contentious soft fork resolves cleanly or fractures into a persistent chain split is difficult to predict analytically but can be explored systematically through controlled experimentation. This paper applies Scenario Discovery, an ensemble-simulation methodology using the Patient Rule Induction Method (PRIM) to identify the configuration regions that produce contentious fork outcomes. Using Warnet to run real \texttt{bitcoind} nodes across 1{,}330 valid scenarios spanning economic weight distribution, mining-pool commitment, pool ideological tolerance, hashrate distribution, and difficulty-retarget regime, we discover the parameter thresholds at which fork behavior transitions from clean resolution to contested split. We introduce a dual-layer analysis that integrates technical consensus metrics with economic-topology metadata, and show that the economic weight distribution across a partition, not hashrate majority, is the primary determinant of resolution under Bitcoin's operational retarget interval and moderate price divergence. Three findings follow: an economic-support floor ($\sim$0.45--0.50) and override ceiling ($\sim$0.78--0.82) that bound the contested space, with an Economic Self-Sustaining Point ($\sim$0.74) between them; a pool-commitment ``flip-point'' ($\sim$0.214 of committed hashrate) at which committing the largest pool to the upgrading chain paradoxically reverses the outcome; and a two-layer outcome structure in which hashrate resolution and economic adoption are governed by different parameters and can resolve independently. Individual user nodes show no detectable influence on fork outcomes under the modeled economic weightings. We close with three monitoring questions, answerable from publicly observable data, that translate these thresholds into operational guidance.
\end{abstract}

}

\maketitle
\IEEEdisplaynontitleabstractindextext
\IEEEpeerreviewmaketitle

\section{Introduction}
\IEEEPARstart{B}{itcoin's} decentralized consensus is usually described in terms of its protocol rules, proof-of-work, heaviest-chain selection, and deterministic block validation. But the operational reality of the network introduces a layer of complexity that pure protocol analysis cannot capture. The network comprises tens of thousands of independently operated nodes, each configured according to its operator's priorities and, more consequentially for governance, each carrying a different amount of economic weight. When a consensus change is proposed, whether the network converges on a single chain or fractures into a durable split is not decided by the protocol specification alone; it is decided by the collective behavior of miners, exchanges, custodians, payment processors, and users, whose incentives do not always align.

As Bitcoin approaches the scale and systemic importance of critical infrastructure with institutional custody, exchange-traded products, and nation-state reserve holdings now material fractions of circulating supply, the question of how contentious upgrades resolve becomes a question of infrastructure risk. De Wolf~\cite{dewolf} maps Bitcoin's protocol and node layers to the IEC~62443 industrial-control-systems security framework, demonstrating that Bitcoin's operational structure closely parallels the architecture of power grids and pipelines. If Bitcoin is critical infrastructure, its failure modes should be measurable.

The Bitcoin community has long recognized, informally, that different classes of network participant hold different kinds of governance power. The BCAP actor framework~\cite{bcap} enumerates the participant classes whose support or opposition shapes the risk of a chain split among them developers, miners, economic nodes, and full-node users, but describes their governance powers only qualitatively. What has been missing is a quantitative account: which of these actors actually determines fork outcomes, under what conditions, and at what thresholds. Existing approaches focus either on protocol-level formal verification or on post-hoc analysis of historical incidents; neither answers the operational question of what specific combinations of economic and mining conditions push a network from clean resolution into a contested split.

This paper addresses that gap directly. We treat contentious fork resolution as an outcome to be \emph{discovered} rather than assumed, applying Scenario Discovery, a methodology from the "decision making under deep uncertainty" tradition to a large ensemble of controlled network simulations. We run real Bitcoin Core nodes under the Warnet framework~\cite{warnet} across a systematically sampled parameter space, define our outcome of interest as the emergence of contentious or persistent forks, and use PRIM~\cite{prim} together with random-forest and logistic analysis to converge on the interpretable parameter regions that produce those outcomes.

\subsection{Research Questions}
\textbf{RQ1.} Which network participants and configuration parameters causally determine the resolution of a contentious soft fork, and can quantitative thresholds be identified that separate clean resolution from persistent split?

\textbf{RQ2.} How does the economic topology of a network-the distribution of custody holdings and transaction volume across a partition-affect the severity and resolution of consensus divergence, relative to the distribution of hashrate?

\textbf{RQ3.} Can a systematic, simulation-based Scenario Discovery framework provide operationally useful, real-time monitoring guidance for node operators, protocol developers, and governance actors during a live contentious fork?

\subsection{Contributions}
This paper makes five contributions. \emph{First}, we present an ensemble-simulation framework built on Warnet that runs real \texttt{bitcoind} nodes and enables systematic Scenario Discovery across the governance parameter space of a contentious fork. \emph{Second}, we introduce a dual-layer analysis model that integrates technical consensus metrics with economic-topology metadata. \emph{Third}, we provide a structural quantification of the BCAP actor classes: an economic-support floor ($\sim$0.45--0.50) and override ceiling ($\sim$0.78--0.82), an Economic Self-Sustaining Point ($\sim$0.74), a pool-commitment flip-point ($\sim$0.214 of committed hashrate), and an ideology $\times$ loss-tolerance product ($\sim$0.16--0.20) governing defender resilience. \emph{Fourth}, we establish that fork outcomes resolve on two independent layers, hashrate and economic adoption, governed by different parameters, and that individual user nodes have no detectable influence on outcomes at the modeled economic weightings. \emph{Fifth}, we translate these findings into three monitoring questions answerable from publicly observable data.

\section{Background and Related Work}

\subsection{Consensus, Forks, and Actor Classes}
Bitcoin's consensus combines proof-of-work mining with deterministic validation rules that all full nodes independently enforce. In principle all nodes following the same rules converge on the chain with the greatest cumulative proof-of-work; in practice, nodes run different software versions and represent very different economic stakes. A \emph{hard fork} loosens rules so that new-rule blocks are rejected by old nodes; without near-universal adoption it produces a persistent split. A \emph{soft fork} tightens rules, so new-rule blocks remain valid to old nodes but not vice versa. Soft forks are designed to be safer but are not safe by default: if adoption is incomplete, the two node sets can still end up on different chains, and markets form around each chain's token, feeding price divergence back into miner incentives. The 2017 Bitcoin Cash split is the most empirically useful case for the present work: a minority chain launched with roughly 5--10\% of hashrate and sustained a divergent market for an extended period, illustrating the interaction of token valuation, hashrate deficit, and difficulty adjustment that our simulation reproduces.

The BCAP framework~\cite{bcap} identifies several participant classes whose support shapes a fork's outcome, of which four are central to soft-fork resolution: developers, who write and vet code; miners, who produce blocks; economic nodes, exchanges, payment processors, custody providers, and large merchants whose custody and settlement decisions define which chain the market treats as ``Bitcoin''; and full-node users, who validate independently but individually carry little economic weight. BCAP describes these powers qualitatively. This paper takes the BCAP taxonomy as its organizing lens and supplies the missing quantification for post fork event consensus. A complementary treatment of the same classes from a security-threat-model perspective is provided by de Wolf~\cite{dewolf}, who identifies mining-pool concentration, node-level eclipse and isolation attacks, and development-governance centralization as the principal threat surfaces corresponding to the BCAP classes.

Two activation strategies frame the governance tension the simulation probes. In a Miner-Activated Soft Fork (MASF), miners signal readiness through block version bits and the rule activates once a hashrate threshold is met. In a User-Activated Soft Fork (UASF), node operators enforce the new rules on a flag day regardless of miner signaling, on the theory that blocks the economic network refuses to accept are worth less to miners. MASF locates the activation lever with miners; UASF locates it with users and the economic infrastructure they are presumed to represent. Which actor actually holds decisive power is an empirical question this paper answers.

\subsection{Scenario Discovery and PRIM}
Scenario Discovery is a methodology from the decision-making-under-deep-uncertainty literature for identifying the conditions under which a system produces an outcome of concern. Rather than predicting a single future, the analyst runs a large ensemble of simulations across uncertain inputs, defines a binary outcome of interest, and searches for the low-dimensional, interpretable region of the input space that best predicts it. The workhorse algorithm is PRIM~\cite{prim,bryant}, which iteratively ``peels'' ranges off the parameter space to isolate axis-aligned boxes within which the outcome density is maximized, yielding human-readable bounds of the form ``contentious outcomes concentrate when parameter $X$ lies below threshold $t$.'' Here the objective is the emergence of contentious or persistent forks; PRIM, supported by random-forest feature-importance analysis and logistic regression with interaction terms, is used to discover which parameters drive that objective and where their thresholds lie.

\subsection{Modeling and Simulation of Blockchain Systems}
Using modeling and simulation to study decentralized blockchain systems is now an established methodology, motivated by the impracticality of experimenting on a production network. This body of work has concentrated on security and performance. The most direct antecedent is Gervais et al.~\cite{gervais}, who introduced a quantitative simulation framework to compare the security and performance of proof-of-work blockchains under varying consensus and network parameters and used it to derive optimal adversarial strategies. Their premise, that security cannot be read off the protocol in isolation but depends on network-layer parameters that must be simulated is one this paper shares and extends from the security layer to the governance layer. General-purpose simulators include BlockSim~\cite{blocksim}, a discrete-event framework organized into network, consensus, and incentive layers, and SimBlock~\cite{simblock}, a network-focused simulator modeling on the order of ten thousand nodes. Foytik et al.~\cite{foytik20} used and modified an NS3-based simulator that captures real network-layer dynamics (latency variance, topology heterogeneity, bandwidth constraints), establishing that consensus behavior is sensitive to network environment in ways fixed-topology models miss. Neudecker et al.~\cite{neudecker} modeled attacks on the Bitcoin peer-to-peer network, and Decker and Wattenhofer~\cite{decker} measured information propagation, linking propagation delay to the stale-block rate.

A complementary question is how to measure decentralization itself. Gochhayat et al.~\cite{gochhayat} develop a multi-dimensional decentrality index whose components, node distribution, hash-power concentration, wealth distribution, and network connectivity, collectively determine how decentralized a network is. The present work operationalizes one of those dimensions, the economic-weight distribution across the partition as the primary governance variable, and the results confirm it dominates the others. On incentive stability, Eyal and Sirer~\cite{eyal} showed that honest-majority mining is not sufficient for incentive compatibility, and Carlsten et al.~\cite{carlsten} extended this to the fee-only regime; Cong et al.~\cite{cong} characterize equilibrium pool formation and miner migration; and Bonneau et al.~\cite{bonneau} systematize the open problems, including the dynamics of miners' pool choice and how stability is affected as block rewards decline, problems this work bears on directly.

Across this literature the metrics are predominantly technical, stale-block and fork rates, throughput, propagation latency, resistance to double-spending, and actors are represented at the network and consensus layers. Two gaps follow. First, prior frameworks abstract the node as a behavioral model; this work runs real \texttt{bitcoind} instances so that validation, chain selection, and propagation are executed by production consensus code. Second, prior simulators do not represent the \emph{economic topology}, the custody and settlement weight of exchanges, custodians, and payment processors, that determines which chain a contested fork resolves to. By adding an economic-adoption layer and coupling the ensemble to Scenario Discovery, this work extends blockchain modeling from how a chain can be attacked to how a contentious fork resolves.

\section{Methodology}

\subsection{Simulation Platform}
Warnet~\cite{warnet} is a Kubernetes-based framework for running controlled Bitcoin network experiments. It launches real \texttt{bitcoind} nodes, not mocks, in cluster pods, connects them according to a declared topology, and lets a Python scenario script coordinate activity across them. A commander process holds RPC handles to every node and drives the simulation: mining blocks, querying state, and applying the behavioral logic below. Because the nodes are genuine Bitcoin Core instances, block validation, chain selection, and propagation follow real consensus code; the scenario layer governs only the economic and strategic behavior of the actors.

\subsection{Network Model and Economic Calibration}
\label{sec:econcalib}
The network models the Bitcoin ecosystem at approximately 1:400 scale, calibrated against public network-measurement, mining-distribution, and exchange proof-of-reserve data as of February~2026. A primary 60-node network represents mining pools, economic nodes, and user nodes individually; a memory-efficient 25-node network preserves the mining layer exactly while consolidating economic and user nodes into weighted aggregate cohorts for high-throughput sampling. All regime-comparison and boundary-fitting analyses use the full 60-node network. The full network comprises eight mining pools (four per partition), 24 economic nodes spanning five roles, and 28 user nodes. Calibration targets three quantities. Hashrate is distributed to match top pools by blocks found (named pools $\approx$86\%; user-node solo mining $\approx$12\%; measurement variance $\approx$2\%). Economic weight is custody-driven: a node's economic weight is $0.7\times\text{custody}_{\text{btc}} + 0.3\times\text{volume}_{\text{btc}}$. Under this calibration, economic nodes hold roughly 96.5\% of total economic weight, with four major-exchange nodes alone controlling about two-thirds of it as a structural concentration central to the results.

A soft fork tightens rules, so the propagation relationship is asymmetric. The network is partitioned at startup into a strict island (v27) and a permissive island (v26); nodes communicate freely within an island but not across. Because stricter-rule blocks are valid under looser rules but not vice versa, v26 nodes accept v27 blocks while v27 nodes reject v26 blocks: a v27 block is submitted to a designated v26 bridge node and propagates through the permissive island by ordinary P2P relay. If a reunion event is enabled, the partitions reconnect and Bitcoin's heaviest-chainwork rule resolves the dispute. The initial division of the network between the two islands is set by two scenario parameters, \texttt{hashrate\_split} and \texttt{economic\_split}, each expressed as the fraction of total hashrate (respectively, economic weight) assigned to the v27 partition at scenario start e.g.\ $\texttt{economic\_split}=0.60$ means 60\% of economic weight begins on v27 and 40\% on v26. All split values reported throughout this paper are v27 shares in this sense; a value below 0.50 indicates v27 starts as the minority partition on that dimension.

\subsection{Actor Behavioral Models}
\textbf{Mining pools.} Eight pools are modeled individually, each characterized by hashrate share, an ideological fork preference (v27, v26, or neutral), an ideology strength, a profitability threshold, and a maximum acceptable revenue loss. Expected hourly profit on a fork is evaluated under a neutral 50/50 hashrate counterfactual to prevent a feedback loop in which a fork's currently low hashrate makes it appear unprofitable. A pool holds its preferred fork while its accumulated loss stays within $\text{ideology\_strength}\times\text{max\_loss\_pct}$, switching to the rational choice (a ``forced switch'') once loss exceeds tolerance. Committed pools hold across substantial profitability gaps and, once forced to switch, rarely switch back absent an external price reversal. Three archetypes emerge: committed pools, neutral profit-maximizers, and moderate ``swing'' pools. The share of non-neutral (ideologically committed) pool hashrate that prefers v27 \texttt{pool\_committed\_split} (C) is a v27 share in the same sense as \texttt{hashrate\_split} and \texttt{economic\_split} above: $\texttt{C}=0.30$ means 30\% of committed hashrate prefers v27 and 70\% prefers v26. \texttt{pool\_neutral\_pct} separately sets what fraction of total pool hashrate carries no ideological preference at all.

\textbf{Economic and user nodes.} Economic nodes represent holders and transactors whose custody and settlement behavior determines market demand for each chain's token; each uses the same ideology-override logic followed by an inertia check, with a node's contribution to a chain's economic weight equal to its custody weight. User nodes represent individual participants, retail holders, power users, solo miners, collectively carrying about 12\% of hashrate and a small share of economic weight; they use the same decision pipeline with higher ideology and inertia.

\subsection{Oracles}
\textbf{Price.} Price diverges only after a fork becomes sustained (a minimum split depth). Once sustained, each fork's price is a weighted combination of three normalized factors:
\begin{equation}
\text{price} = \text{base}\times(0.3\,c_{\text{chain}} + 0.5\,c_{\text{econ}} + 0.2\,c_{\text{hash}}),
\end{equation}
capped at $\pm$20\% divergence per chain from the pre-fork base price (so the ratio between the two chains' prices can reach $\sim$45\%). The dominance of the economic coefficient reflects Bitcoin's custody-driven valuation and is consistent with wavelet-coherence evidence that custody and volume signals lead price relative to hashrate~\cite{kristoufek}; the coordination tipping point of the ``blockchain folk theorem''~\cite{biais} and marginal-cost-of-production convergence~\cite{hayes} imply the linear mapping and symmetric cap conservatively \emph{understate} minority-chain collapse.
\textbf{Difficulty and chainwork.} After a fork, both chains inherit pre-fork difficulty; the minority chain produces blocks more slowly until it retargets. A difficulty oracle models per-fork block production probabilistically, retargets every retarget-interval blocks, and tracks cumulative chainwork which is the quantity that determines a reunion winner. \textbf{Fees.} A slower minority chain accumulates mempool volume, raising fee rates and partially compensating its miners; this effect is small relative to the price differential in most scenarios.

\subsection{Simulation Loop, Noise Floor, and Discovery Procedure}
The scenario runs a tick-based loop: probabilistic block production on each fork, cross-partition propagation of v27 blocks, state collection, and periodic price/decision updates. Block production is the only stochastic element; all decision processes are deterministic threshold functions, so outcome variance across identical runs reflects mining stochasticity alone. A dedicated baseline sweep (\texttt{balanced\_baseline}; 27 of 30 seeded runs) at symmetric 47\%/47\% starting hashrate with no committed ideology measured this noise floor at $\sigma=3.3\%$ block-share variance, with zero cascades and zero reorgs. The systematic effects reported below produce shifts of 15--50\%, an order of magnitude above baseline.

Scenarios are generated by two complementary methods. Latin Hypercube Sampling (LHS) spreads scenarios uniformly across the parameter space for an unbiased view before any structural hypothesis is formed; targeted grid sweeps then vary one or two parameters while holding others fixed to confirm or falsify causal claims and map thresholds. Two retarget regimes are tested: a 144-block interval ($\approx$1~day, accelerated) and a 2016-block interval ($\approx$14~days, matching Bitcoin mainnet). Three analysis methods are applied: PRIM isolates axis-aligned boxes maximizing outcome uncertainty or a composite contentiousness score (combining reorg count, reorg depth, and cascade timing into a single disruption index); random-forest classification predicts the binary outcome and reports feature importance and out-of-bag (OOB) accuracy; logistic regression with interaction terms captures interactions not visible in single-parameter analysis. The principal modeling assumptions, 50/50 profitability counterfactual, $\pm$20\% price cap, static custody-based economic weight, fixed subsidy/cost within a run, and effectively irreversible forced-switch commitment, are revisited in Section~\ref{sec:limits}.

\section{Results}
This section reports findings from 1{,}374 executed scenarios (1{,}330 valid, 96.8\%; the 24 excluded runs failed startup or configuration validation due to a role-name parameter bug in early lite-network sweeps) across 21 sweep configurations, organized in discovery order: parameter causality, active causal parameters and thresholds, regime comparison, and governance implications. Of the 1{,}330 valid scenarios, 268 are 144-block full-network runs and 590 are 2016-block full-network runs (Phase~2, $n{=}298$, plus Phase~3, $n{=}292$); the remaining 472 are 25-node reduced-network runs used only for high-throughput LHS sampling. All regime-comparison and boundary-fitting analyses use full-network scenarios only.

\subsection{Separating Signal from Noise}
The \texttt{balanced\_baseline} sweep establishes 3.3\% as the noise floor: any block-share shift exceeding this indicates active parameter influence. Initial LHS sweeps identified \texttt{hashrate\_split} as the apparent dominant predictor (Spearman $r=+0.83$), but targeted isolation revealed this to be a sampling artifact: high-hashrate scenarios co-occurred with pool configurations independently favorable to v27. When \texttt{hashrate\_split} was varied in isolation across a $6\times7$ grid from 0.15 to 0.65 with all other parameters at medians (144-block), outcomes were \emph{identical across all six hashrate levels at every economic level} (Table~\ref{tab:hash}): fork outcomes are determined entirely by \texttt{economic\_split}, with hashrate contributing zero independent effect. The mechanism is the \emph{Difficulty Adjustment Survival Window} (Section~\ref{sec:window}): the minority chain's difficulty adjusts downward as blocks slow, equalizing block-production rates regardless of starting split before the economic cascade resolves.

\begin{table}[!t]
\caption{Fork outcomes across \texttt{hashrate\_split}$\times$\texttt{economic\_split} (144-block). Identical columns across hashrate levels confirm non-causality.}
\label{tab:hash}
\centering\small
\setlength{\tabcolsep}{4pt}
\begin{tabular}{lccccccc}
\toprule
\textbf{hash}$\backslash$\textbf{econ} & 0.35 & 0.45 & 0.50 & 0.55 & 0.60 & 0.70 & 0.82\\
\midrule
0.15 & \rcell v26 & \gcell v27 & \gcell v27 & \gcell v27 & \rcell v26 & \rcell v26 & \gcell v27\\
0.25 & \rcell v26 & \gcell v27 & \gcell v27 & \gcell v27 & \rcell v26 & \rcell v26 & \gcell v27\\
0.35 & \rcell v26 & \gcell v27 & \gcell v27 & \gcell v27 & \rcell v26 & \rcell v26 & \gcell v27\\
0.45 & \rcell v26 & \gcell v27 & \gcell v27 & \gcell v27 & \rcell v26 & \rcell v26 & \gcell v27\\
0.55 & \rcell v26 & \gcell v27 & \gcell v27 & \gcell v27 & \rcell v26 & \rcell v26 & \gcell v27\\
0.65 & \rcell v26 & \gcell v27 & \gcell v27 & \gcell v27 & \rcell v26 & \rcell v26 & \gcell v27\\
\bottomrule
\end{tabular}
\end{table}

A dedicated 2016-block verification ($n{=}18$) confirms and qualifies this: at $\text{econ}=0.60$ and $0.70$ all cells produce v27 wins regardless of hashrate, but at economic parity ($\text{econ}=0.50$) hashrate becomes \emph{conditionally} causal, exhibiting non-monotonic behavior because the wider survival window lets intermediate v27 hashrate build a v26 block-length lead that pushes the largest pool past its loss tolerance before difficulty relief arrives. This conditional causality is regime-dependent and qualifies rather than reverses the non-causality result: hashrate is non-causal at $\text{econ}\ge0.60$, which covers the realistic range of contested forks with meaningful economic support on either side.

Eight further parameters were confirmed non-causal (Table~\ref{tab:noncausal}). Of particular note is the \emph{exact} null result for all three user-behavior parameters: no output metric varied across any of the 36 scenarios in the user sweep. User nodes collectively represent a 2{,}197:1 economic-weight disadvantage relative to institutional actors and cannot exert pricing power independently of the exchanges and custodians that set miner revenue. This ratio characterizes non-cooperative user nodes, individual holders whose custody and transaction volume are modeled as small and largely static (Section~\ref{sec:econcalib}); it does not bound what a coordinated user population could achieve. A bloc of users that pooled and actively represented its aggregate holdings and transaction volume behaving, in economic-weight terms, like a business or payment processor rather than individual holders would satisfy the model's own custody-and-volume-weighted definition of economic weight and would functionally reclassify as an economic node, subject to the economic-node dynamics described throughout this paper rather than the null result reported here. After elimination, three active causal parameters remain: \texttt{economic\_split}, \texttt{pool\_committed\_split}, and the \texttt{pool\_ideology\_strength}\,$\times$\,\texttt{pool\_max\_loss\_pct} interaction.

\begin{table}[!t]
\caption{Parameters confirmed non-causal through targeted sweeps.}
\label{tab:noncausal}
\centering\footnotesize
\setlength{\tabcolsep}{3.5pt}
\renewcommand{\arraystretch}{1.15}
\begin{tabular}{@{}p{2.9cm}p{0.8cm}p{3.65cm}@{}}
\toprule
\textbf{Parameter} & \textbf{Fixed} & \textbf{Evidence}\\
\midrule
\texttt{hashrate\_\allowbreak split} & 0.25 & Zero effect over 0.15--0.65 at econ $\ge$ 0.60; conditionally causal only at econ $=$ 0.50 (2016-block).\\
\texttt{pool\_\allowbreak neutral\_\allowbreak pct} & 30\% & Controls cascade duration only; outcome unchanged 10\%--50\%.\\
\texttt{econ\_\allowbreak inertia} & 0.17 & No effect on full 60-node network.\\
\texttt{econ\_\allowbreak switching\_\allowbreak thr.} & 0.14 & No effect on full 60-node network.\\
\texttt{user\_\allowbreak ideology\_\allowbreak str.} & 0.49 & Exact null: Spearman $r=0.000$.\\
\texttt{user\_\allowbreak switching\_\allowbreak thr.} & 0.12 & Exact null: Spearman $r=0.000$.\\
\texttt{user\_\allowbreak nodes/\allowbreak part.} & 6 & Exact null: Spearman $r=0.000$.\\
\texttt{pool\_\allowbreak profit\_\allowbreak thr.} & 0.16 & Importance separation 0.011; below detection.\\
\texttt{solo\_\allowbreak miner\_\allowbreak hashrate} & 0.085 & Separation $\approx$ 0; follows profit signal.\\
\bottomrule
\end{tabular}
\end{table}

\subsection{Causal Parameters and the Decision Boundary}
\begin{figure}[!t]
\centering
\includegraphics[width=\linewidth]{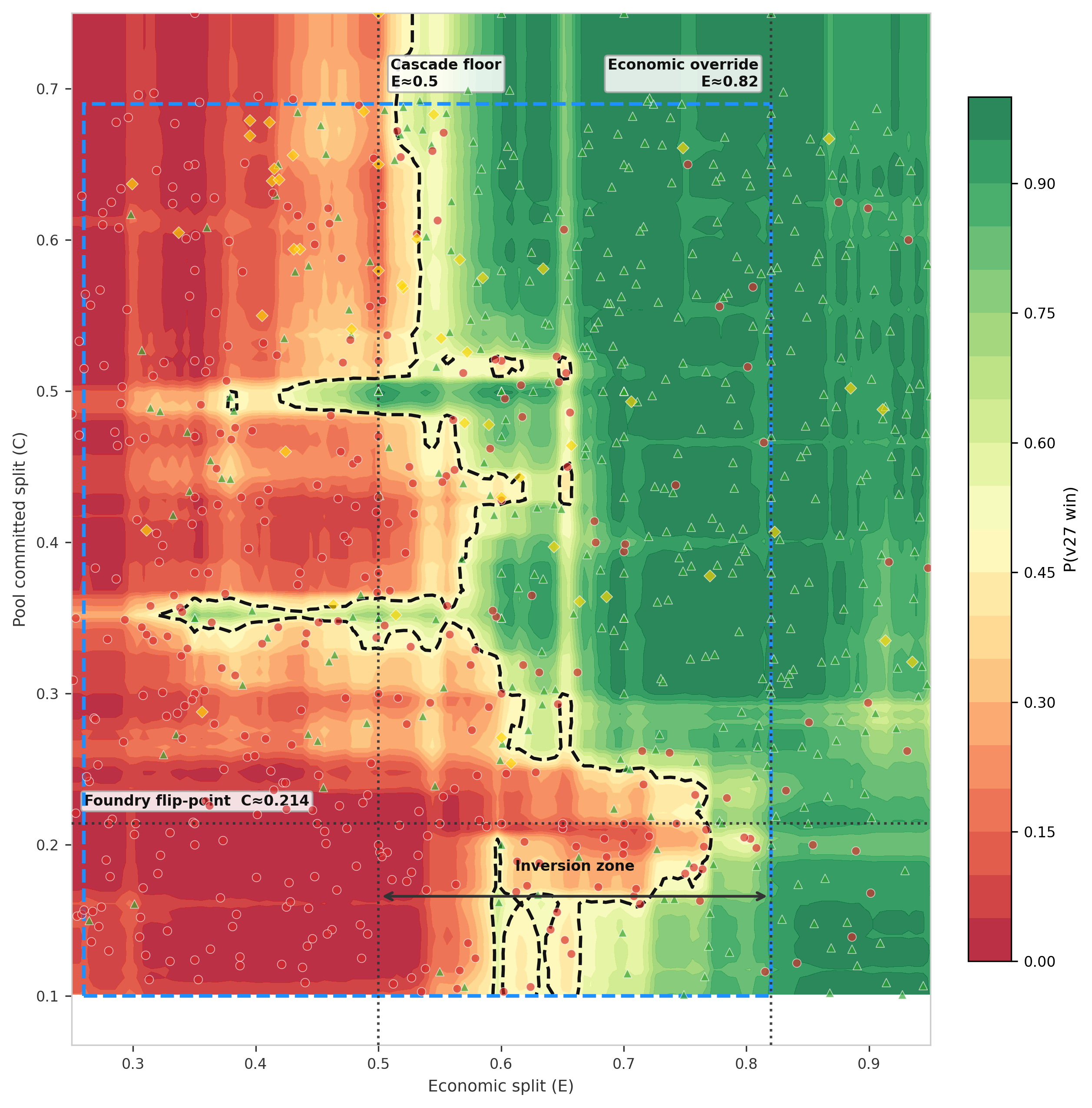}
\caption{Random-forest P(v27~win) decision surface across \texttt{economic\_split} (E) and \texttt{pool\_committed\_split} (C), fitted on all valid full-network 2016-block scenarios ($n{=}837$, 7 sweeps: unbiased LHS plus targeted sweeps that oversample the inversion zone; RF OOB $=$ 81.2\%). Lite-network scenarios are excluded so the figure is on the same full-network basis as the rest of the paper's regime-comparison and boundary-fitting results. Green triangles denote v27-dominant outcomes, red circles v26-dominant, and yellow diamonds contested outcomes; the dashed black line is the $P{=}0.50$ decision contour. Dotted lines mark the structural thresholds: the cascade floor (E\,$\approx$\,0.50), the economic override (E\,$\approx$\,0.82), and the Foundry flip-point (C\,$\approx$\,0.214). The blue box marks the PRIM uncertainty region. Note the inversion within the contested band: at intermediate economic support, crossing the flip-point from C\,$<$\,0.214 to C\,$>$\,0.214 reverses the outcome from v27 to v26.}
\label{fig:boundary}
\end{figure}

\textbf{Economic threshold and inversion zone.} Mapping outcomes across an \texttt{economic\_split}$\times$\texttt{pool\_committed\_split} grid (Table~\ref{tab:inversion}, Fig.~\ref{fig:boundary}) reveals a non-monotonic boundary with three regimes. In the weak-economics regime ($\text{econ}\le0.45$), no cascade is possible and v26 wins across all pool configurations. In the strong-economics regime ($\text{econ}\ge0.82$), the price signal breaks even strongly committed v26 pools and v27 wins universally. The intermediate regime exhibits an \emph{inversion}: outcomes are non-monotonic in \texttt{pool\_committed\_split}, with v27 winning only at the lowest committed level and losing at all higher levels, a counter-intuitive result explained by the Foundry flip-point.

\begin{table}[!t]
\caption{Fork outcomes across \texttt{economic\_split}$\times$\texttt{pool\_committed\_split}. Note the inversion at econ $=0.60$--$0.70$ where the effect of pool commitment reverses sign. $\dagger$: partial cascade, v26 retains dominance.}
\label{tab:inversion}
\centering\small
\setlength{\tabcolsep}{2.6pt}
\begin{tabular}{lccccccccc}
\toprule
\textbf{econ}$\backslash$\textbf{C} & .20 & .30 & .38 & .43 & .47 & .52 & .58 & .65 & .75\\
\midrule
0.35 & \rcell v26 & \rcell v26 & \rcell v26 & \rcell v26 & \rcell v26 & \rcell v26 & \rcell v26 & \rcell v26 & \rcell v26\\
0.50 & \rcell v26 & \gcell v27 & \gcell v27 & \gcell v27 & \gcell v27 & \gcell v27 & \gcell v27 & \gcell v27 & \gcell v27\\
0.60 & \gcell v27 & \rcell v26 & \rcell v26 & \rcell v26 & \rcell v26 & \rcell v26 & \rcell v26 & \rcell v26 & \rcell v26\\
0.70 & \gcell v27 & \rcell v26$\dagger$ & \rcell v26$\dagger$ & \rcell v26$\dagger$ & \rcell v26$\dagger$ & \rcell v26 & \rcell v26 & \rcell v26 & \rcell v26\\
0.82 & \gcell v27 & \gcell v27 & \gcell v27 & \gcell v27 & \gcell v27 & \gcell v27 & \gcell v27 & \gcell v27 & \gcell v27\\
\bottomrule
\end{tabular}
\end{table}

\textbf{The Foundry flip-point.} The inversion is caused by a structural feature of the pool distribution. The largest pool, Foundry USA ($\approx$30\% of total hashrate, the largest pool in this calibration), crosses from v26-preferring to v27-preferring assignment at $\texttt{pool\_committed\_split}\times0.70>0.15$, i.e.\ a threshold of $\approx$0.214. Below it, Foundry is assigned v26 but at 60--70\% economic support the v27 premium exceeds its loss tolerance, trapping it economically and forcing it to v27, providing the committed anchor that enables the cascade. Above it, Foundry shifts to v27 by conviction, but the reassignment simultaneously hardens the opposing v26 bloc (AntPool $\approx$18\% and F2Pool $\approx$15\%, together $\approx$33\%), which is now sufficient to resist the cascade at moderate economic support. A four-percentage-point shift in \texttt{pool\_committed\_split} from 0.20 to 0.30, reverses the outcome entirely. The flip-point is confirmed by unbiased LHS at 2016-block ($n{=}64$): all 12 v26-dominant cases have $\texttt{committed}\le0.246$ and all 52 v27-dominant cases have $\texttt{committed}\ge0.260$, with a clean gap at 0.247--0.259 confirming a structural threshold rather than a smooth gradient. A complementary robustness check varies \emph{which} pools compose the committed set at a fixed nominal $C$, rather than the deterministic hashrate-order assignment used above: across the \texttt{pool\_composition\_arm\_a} sweep (168 scenarios), v27 wins in 0 of 24 runs at nominal $C=0.214$, because random compositions at that target rarely happen to include a pool as large as Foundry, leaving realized committed hashrate well below the nominal value ($\approx$0.146 on average). This is consistent with, rather than contrary to, the mechanism above: the operative quantity is \emph{realized committed hashrate} crossing $\approx$0.214--0.26, not the identity of any specific pool, Foundry is simply the largest pool in the modeled landscape, so it is the pool whose assignment happens to move realized hashrate across that threshold in the main sweeps. The ``Foundry flip-point'' is thus shorthand for a major-pool flip-point: whichever single pool is large enough to shift realized committed hashrate across the threshold produces the same reversal, and in this calibration that pool is Foundry.

\textbf{Pool ideology threshold.} A second threshold determines whether the cascade \emph{completes}. The interaction of \texttt{pool\_ideology\_strength} and \texttt{pool\_max\_loss\_pct}, whose product defines a pool's maximum acceptable loss, gates whether committed v26 pools capitulate or hold. Mapped on the full network at $\text{econ}=0.78$, a diagonal threshold appears: committed v26 pools survive when ideology$\times$max\_loss is $\approx$0.16--0.20; below this, pools capitulate regardless of individual ideology or loss tolerance. Neither parameter is sufficient alone, the product is the operative quantity. Above $\text{econ}\approx0.82$, a dedicated override sweep ($n{=}27$) produced v27-dominant outcomes in all cases: the price signal forces capitulation regardless of the product. The product nonetheless continues to govern cascade \emph{timing}: at high ideology the cascade can take over 15$\times$ longer than at low ideology at the same economic level, so the \emph{duration} of a contested fork may be governed by pool ideology even when the outcome is not. Between the inversion zone and the override ceiling lies an Economic Self-Sustaining Point (ESP; $\approx$0.74): above it, economic momentum on the upgrading chain becomes self-reinforcing neutral pools and economic nodes continue migrating without further external pressure though committed opposition can still delay resolution.

\begin{table}[!t]
\caption{Consolidated threshold estimates. All ratings High-confidence, from full-network targeted sweeps.}
\label{tab:thresholds}
\centering\footnotesize
\setlength{\tabcolsep}{3.5pt}
\renewcommand{\arraystretch}{1.2}
\begin{tabular}{@{}p{2.85cm}p{1.35cm}p{3.5cm}@{}}
\toprule
\textbf{Parameter} & \textbf{Threshold} & \textbf{Interpretation}\\
\midrule
\texttt{economic\_\allowbreak split} (cascade floor) & $\sim$0.45--0.50 & Below this, no cascade; v26 wins regardless of pool configuration.\\
\texttt{economic\_\allowbreak split} (inversion onset) & $\sim$0.55--0.60 & Effect of \texttt{pool\_\allowbreak committed\_\allowbreak split} reverses sign.\\
\texttt{economic\_\allowbreak split} (ESP) & $\sim$0.74 & Economic momentum on the upgrading chain becomes self-reinforcing: neutral pools and economic nodes keep migrating without further external push.\\
\texttt{economic\_\allowbreak split} (economic override) & $\sim$0.78--0.82 & Economic signal overrides all pool configuration; v27 wins universally regardless of pool ideology.\\
\texttt{pool\_\allowbreak committed\_\allowbreak split} (Foundry flip) & $\sim$0.214 & Reassigns Foundry ($\approx$30\% hash) and inverts outcomes at econ $=0.60$--$0.70$; hard gap 0.247--0.259.\\
ideology $\times$ max\_\allowbreak loss (product) & $\sim$0.16--0.20 & Below: committed pools capitulate; above: hold (at econ $=0.78$). Vanishes at econ $\ge0.82$.\\
\bottomrule
\end{tabular}
\end{table}

For a governance actor monitoring a fork, these thresholds (Table~\ref{tab:thresholds}) reduce to three questions: is \texttt{economic\_split} above $\approx$0.50 (else the upgrading fork cannot win); is \texttt{pool\_committed\_split} above or below $\approx$0.214 (which side is the largest pool on, and by entrapment or conviction); and is \texttt{economic\_split} above $\approx$0.82 (above which pool structure is irrelevant to the outcome, though it still governs timing).

\subsection{Regime Comparison: The Causal Rank Reversal}
\label{sec:window}
The dominant causal parameter changes entirely with the retarget interval. % TODO (before TCSS): report RF importances with confidence intervals / repeated seeds and a held-out test, not OOB alone; report logistic-regression coefficients and PRIM box coverage/density from the technical report.
Random-forest classification fitted separately on the 144-block ($n{=}268$) and the Phase-2 2016-block ($n{=}298$) full-network datasets, using the same four active parameters, shows a rank reversal (Table~\ref{tab:rf}). At 144-block, \texttt{economic\_split} accounts for 77.2\% of predictive importance about four times \texttt{pool\_committed\_split} (11.3\%). At 2016-block, \texttt{pool\_committed\_split} displaces it entirely (52.8\% vs.\ 20.2\%). The two parameters swap rank, and neither holds even a plurality in the other's regime. The 2016-block regime is Bitcoin's operationally relevant one, making pool commitment the primary governance result. Secondary observations: 2016-block outcomes are more predictable (OOB 83.2\% vs.\ 80.0\%), and \texttt{pool\_max\_loss\_pct} rises to a meaningful secondary factor (17.1\%) because committed pools must sustain losses through an entire epoch.

\begin{table}[!t]
\caption{Random-forest feature importance by retarget regime.}
\label{tab:rf}
\centering\small
\setlength{\tabcolsep}{5pt}
\begin{tabular}{@{}p{2.5cm}ccc@{}}
\toprule
\textbf{Parameter} & \textbf{144-block} & \textbf{2016-block} & \textbf{Rank}\\
\midrule
\texttt{economic\_\allowbreak split} & \gcell 77.2\% & 20.2\% & \#1\,$\rightarrow$\,\#2\\
\texttt{pool\_\allowbreak committed\_\allowbreak split} & 11.3\% & \gcell 52.8\% & \#2\,$\rightarrow$\,\#1\\
\texttt{pool\_\allowbreak max\_\allowbreak loss\_\allowbreak pct} & 5.5\% & 17.1\% & \#4\,$\rightarrow$\,\#3\\
\texttt{pool\_\allowbreak ideology\_\allowbreak strength} & 6.0\% & 9.9\% & \#3\,$\rightarrow$\,\#4\\
\midrule
RF OOB accuracy & 80.0\% & 83.2\% & \\
\bottomrule
\end{tabular}
\end{table}

The reversal follows from the \emph{Difficulty Adjustment Survival Window}: the period between fork inception and the minority chain's first difficulty retarget, during which it must absorb its hashrate disadvantage without relief. At 144-block the window spans $\approx$24 hours; the economic price signal operates on the same timescale, so the cascade resolves before the retarget fires and economic alignment is the binding constraint. At 2016-block the window spans $\approx$14 days; the price signal still fires within hours to days, but committed pool structure must hold through weeks of sustained losses before relief, so whether committed pools can absorb the full epoch, governed by the ideology$\times$max\_loss product is the binding constraint. This mechanism unifies the findings: hashrate is non-causal because the window neutralizes the majority's advantage before the cascade resolves; pool commitment dominates at 2016-block because the long epoch forces committed pools to reveal whether ideological tolerance exceeds loss tolerance. A corollary \emph{hashrate-parity danger window} appears at 2016-block and $\text{econ}=0.50$: intermediate v27 hashrate (35--45\%) is strictly worse for v27 than either low or high hashrate, because the v26 chain builds a block-length lead quickly enough to push the largest pool past its tolerance within one retarget cycle.

Table~\ref{tab:regime} summarizes the two regimes. Under the operationally relevant 2016-block interval the dominant question is not economic majority but which large pools are committed to each side and whether they can sustain losses through a full 14-day epoch. The 2016-block regime is also 1.6$\times$ more contentious on average, the operationally relevant regime is also the more disruptive one.

\begin{table}[!t]
\caption{Regime comparison across measured dimensions.}
\label{tab:regime}
\centering\footnotesize
\setlength{\tabcolsep}{3.5pt}
\renewcommand{\arraystretch}{1.2}
\begin{tabular}{@{}p{2.15cm}p{2.6cm}p{2.7cm}@{}}
\toprule
\textbf{Dimension} & \textbf{144-block} & \textbf{2016-block}\\
\midrule
Dominant param & \texttt{economic\_\allowbreak split} (77.2\%) & \texttt{pool\_\allowbreak committed\_\allowbreak split} (52.8\%)\\
Boundary shape & $\sim$1-D (economic threshold) & 2-D (E$\times$C interaction)\\
OOB accuracy & 80.0\% & 83.2\%\\
Contentiousness & 0.166 & 0.271 (1.6$\times$)\\
Mechanism & Price cascade resolves before retarget & Pool endurance through full epoch\\
Governance Q & ``Which fork has economic majority?'' & ``Which pools sustain losses 14 days?''\\
\bottomrule
\end{tabular}
\end{table}

\subsection{Implications for Fork Assessment}
The standard framing, whether the new version has majority hashrate support addresses a parameter confirmed non-causal at realistic economic levels and actively misleading within the inversion zone. One data point makes this stark: in a scenario starting at 90\% v27 hashrate but only 7\% v27 economic support, v26 wins after 7 reorgs. A fork beginning with 90\% of network hashrate is defeated when 93\% of economic activity remains on the old chain. This scenario lies below the economic floor, so it illustrates rather than proves the general point, but it dramatizes why hashrate majority is not sufficient for fork victory.

Three monitoring questions better track the causal structure, each answerable from public data (Table~\ref{tab:decision}). \emph{Question~1:} what fraction of economically significant activity is committed to each fork? This is estimable from exchange chain-support policies, ETF fork-handling procedures, and on-chain UTXO spending patterns. \emph{Question~2:} which major pools are ideologically committed versus profit-maximizing, and what is each committed pool's switching-cost threshold? Pool positions are partially observable from public signaling and block attribution; the harder quantity the ideology$\times$max\_loss product could be calibrated against observed behavior during past fork events. \emph{Question~3:} has the fork crossed the override threshold, and if not, which side does the largest committed pool occupy? In the intermediate zone the identity and assignment of the largest committed pool is more consequential than aggregate committed share.

\begin{table}[!t]
\caption{Fork-assessment decision structure by economic condition.}
\label{tab:decision}
\centering\footnotesize
\setlength{\tabcolsep}{3.5pt}
\renewcommand{\arraystretch}{1.2}
\begin{tabular}{@{}p{1.9cm}p{5.0cm}@{}}
\toprule
\textbf{Condition} & \textbf{Implication \& primary signal}\\
\midrule
econ $\ge$ 0.82 & Pool ideology irrelevant to outcome; override total. Monitor cascade \emph{timing} via ideology$\times$max\_loss.\\
0.50 $<$ econ $<$ 0.82 & Outcome set by pool commitment. Is the Foundry-class pool economically \emph{trapped} on the upgrading chain, or ideologically \emph{committed} to it?\\
econ $\le$ 0.50 & No cascade possible. Incumbent wins regardless of pool config or hashrate. No further monitoring needed.\\
\bottomrule
\end{tabular}
\end{table}

The model also identifies a consistent structural sequence pool hashrate consolidation \emph{precedes} economic-node migration and a magnitude-based diagnostic robust to timing uncertainty: inter-chain price gaps of 41--47\% (each chain bounded by the $\pm$20\% per-chain cap) indicate the full-switch regime, gaps of 12--18\% the no-switch regime. A fork that resolves at the hashrate level but shows only 12--18\% price divergence is likely in the no-switch regime, economic adoption will not complete regardless of additional time. Finally, whether the minority chain has reached its first retarget epoch marks a qualitative transition: a fork that appears to be dying may stabilize following that retarget, providing a natural on-chain checkpoint for assessment.

\section{Discussion}

\subsection{Commitment Is Not the Same as Leverage}
The central finding is not merely that economic weight matters more than hashrate claimed qualitatively before, but that the causal structure is counterintuitive in a precise, quantifiable way. The Foundry flip-point demonstrates that \emph{increasing} pool commitment to the upgrading fork can reverse the outcome under intermediate economic conditions. The mechanism is game-theoretic and operates through a forced-exit dynamic with a direct analogue in financial markets. Below the flip-point, the largest pool is economically trapped: the v27 premium exceeds its loss tolerance, and accumulated opportunity cost functions like margin pressure on a leveraged position. When the loss crosses tolerance the exit is involuntary and an involuntary switch is cascade-generating in a way voluntary commitment is not, because it signals to neutral pools that a large operator's pain threshold has been breached. Above the flip-point, Foundry holds v27 by conviction, but its departure from the v26 bloc \emph{purifies} that bloc: AntPool and F2Pool, no longer diluted, are genuinely committed defenders whose combined ideology$\times$max\_loss product resists the signal. The upgrade coalition gains a committed ally but loses the economic-pressure mechanism. The governance implication is non-obvious: a governance actor should ask not ``which large pools support us?'' but ``which large pools are trapped on the opposing chain and will be forced to switch?'' and public commitment by a large pool may, under intermediate economic conditions, be counterproductive.

\subsection{The Two-Layer Outcome Structure}
Fork outcomes operate on two independent causal layers governed by different parameters and resolving on different timescales. \emph{Layer 1 (hashrate outcome)} is determined primarily by \texttt{pool\_committed\_split} relative to the flip-point; at 2016-block it is the dominant factor (52.8\% RF importance). \emph{Layer 2 (economic adoption)} is determined primarily by \texttt{pool\_max\_loss\_pct} ($r=-0.417$ with final economic share): even after one chain establishes block-production dominance, this layer determines whether the winning chain achieves full economic-node migration or stabilizes in a persistent split. The layers are largely independent, a fork can resolve cleanly at the hashrate layer while remaining contested at the economic layer, or vice versa. A hashrate-arbitrage effect adds a third temporal dimension: even after Layer~1 resolves, the minority chain's approaching difficulty adjustment creates a brief profitability spike that may temporarily attract neutral hashrate back, making the interval immediately preceding the first difficulty adjustment the highest-risk window for rapid reversal and it is publicly observable from block-production rates. The 2017 Bitcoin Cash fork, which maintained token value through a weeks-long survival window (aided by emergency difficulty-adjustment modifications), illustrates the same three-variable interaction the simulation produces under standard dynamics.

\subsection{The Structural Ceiling on User-Node Influence}
Across all 598 2016-block scenarios (spanning 15 sweep configurations), no parameter configuration was found in which user nodes are near-pivotal. The Scenario-Potential bias ratio of 1.256 only marginally above the 0.975 baseline, well below the 2.0 that would indicate meaningful concentration is the correct output for the economic-weight ratio: user nodes collectively hold a 2{,}197:1 disadvantage against the full economic network. Even the most favorable conceivable configuration cannot produce user-node pivotality, because the economic weight controlled by exchanges, custodians, and payment processors is structurally dominant. This bears directly on UASF theory. The UASF argument that user nodes enforcing new rules make miners' blocks worth less is real, but the simulation locates its leverage precisely: user nodes create economic pressure only insofar as they \emph{are} the economic infrastructure or control it. Individual full-node operators running updated software are not exchanges, custodians, or payment processors and do not set the prices that determine miner revenue. UASF campaigns succeed when they persuade economic actors to enforce new rules as the 2017 SegWit activation involved credible commitments from major economic actors not when they increase the count of individual full nodes. This does not make user nodes irrelevant, they propagate transactions, enforce policy-layer rules, and provide distributed verification, but the specific claim that an individual operator running strict-validation software can shift a contested soft-fork outcome is not supported.

\subsection{Implications for Protocol Developers and Governance Actors}
The three findings suggest a reframing. The conventional questions, ``what fraction of hashrate supports the upgrade?'' and ``how many nodes run the new software?'', are poor predictors. The simulation-grounded questions are: (1) Is \texttt{economic\_split} above $\sim$0.50? Below it, the upgrading fork cannot win. (2) Is it above $\sim$0.82? Above it, the outcome is determined and only disruption remains in question. (3) In the intermediate range, which side is the largest pool on, and is it there by economic entrapment or ideological commitment? (4) How far is the minority chain from its first difficulty adjustment, and what is the current price differential? These are monitorable in real time from exchange pricing, pool block attribution, and block-production rates.

\subsection{Limitations}
\label{sec:limits}
Several limitations bound how these findings apply. The 2{,}197:1 user-to-economic weight ratio reflects a specific calibration; substantially different weight distributions could shift the structural ceiling on user influence. The $\pm$20\% price-divergence cap does not capture the extreme divergence of real events (BCH/BTC, BCH/BSV reached 80--95\% over months); under larger divergences the losing chain's token may collapse before its first retarget, making hashrate suddenly causal in a way the model does not capture. Network topology is static: the model does not capture new actors entering mid-fork, actors changing position, or second-order strategic behavior such as a large exchange announcing support to influence others. Pool switching is a deterministic threshold-crossing; real operators exercise judgment, switching early to signal, late to extract concessions, or splitting hashrate across chains. Finally, the specific threshold values are calibrated to the modeled 2026 pool distribution and price-oracle weights: the mechanisms are general, but a network with a materially different mining or economic landscape would require recalibration.

\subsection{Future Work and a Broader Program}
This soft-fork causality study is the first of three planned contributions in a Warnet-based critical-scenario-discovery program. Within the present study, priority extensions include the sub-10\% hashrate regime (where the survival-window argument predicts a failure boundary for non-causality), uncapped price-divergence dynamics, dynamic pool timing that anticipates the retarget arbitrage window, and operationalization of the three monitoring questions against live chain data. Beyond it, a second study will apply the same PRIM-based methodology to the full space of Bitcoin node configuration, version mix, mempool policy, network policy, and resource constraints, to discover which combinations produce critical network behavior independent of fork governance. A third study will implement fork-validation differences directly in the Bitcoin Core codebase so that consensus divergence emerges from real peer-to-peer messaging rather than commander-layer orchestration, enabling gossip-network structure itself to be varied as an input and connecting these fork-resolution findings to the connectivity--security questions the Erlay~\cite{erlay} and eclipse-attack~\cite{neudecker} literature identifies as open.

\balance
\section{Conclusion}
This paper applied Scenario Discovery to contentious Bitcoin soft forks, running real \texttt{bitcoind} nodes under Warnet across 1{,}330 valid scenarios and using PRIM, random-forest, and logistic analysis to discover the parameter regions that produce contested outcomes. The central result is that fork resolution under Bitcoin's operational 2016-block retarget interval, at moderate price divergence, is governed not by hashrate majority but by the interaction of economic weight distribution and mining-pool commitment structure. An economic-support floor of $\sim$0.45--0.50 and an override ceiling of $\sim$0.78--0.82 bound the contested space, with an Economic Self-Sustaining Point at $\sim$0.74 between them; within that band the outcome turns on a pool-commitment flip-point at $\sim$0.214, where the ideological assignment of the single largest pool reverses the result, so that publicly committing a large pool to the upgrading chain can be counterproductive at intermediate economic conditions. Pool resilience is governed by an ideology$\times$loss-tolerance product threshold of $\sim$0.16--0.20. Fork outcomes resolve on two independent layers, and individual user nodes, at the modeled economic weightings, show no detectable influence on outcomes. These findings supply the structural quantification the BCAP actor framework leaves qualitative, and translate into three monitoring questions answerable in real time from publicly observable data. The mechanisms are general; the specific threshold values are calibrated to the modeled 2026 landscape and would require recalibration for a materially different network.

\section*{Data and Code Availability}
The scenario configurations, the behavioral- and oracle-model implementations, the outcome dataset for all 1{,}330 scenarios, and the PRIM, random-forest, and logistic-regression analysis scripts will be released in a public repository and archived with a persistent DOI (Zenodo) \url{https://doi.org/10.5281/zenodo.21814612}; repository \url{https://github.com/pfoytik/Bitcoin-Fork-Governance-Study}. The Warnet framework on which the experiments run is open source~\cite{warnet}.

\section*{Acknowledgment}
The author thanks the University of Wyoming Bitcoin Research Institute for organizing the workshop held July 13--17, 2026, where an earlier version of this work was presented, and thanks the workshop attendees for their valuable feedback and discussion.

% ============================================================
% TODO BEFORE TCSS SUBMISSION (deferred from the quick-fix pass):
%  1. ROBUSTNESS/SENSITIVITY (highest priority): show the thresholds
%     and the Foundry flip-point survive perturbation of (a) the oracle
%     coefficients in Eq. (1), (b) the economic-weight calibration, and
%     (c) the price cap. Surface the existing price_divergence_sensitivity_2016
%     sweep (n=48, four cap levels) here.
%  2. FIGURES: restore the original data figures from the long draft --
%     the n=64 LHS flip-point separation (0.247-0.259 gap), a survival-window
%     timing plot, and a PRIM peeling/coverage-density plot. Fig. 1 has been
%     regenerated (2026-08-05) as a real RF P(v27 win) surface over n=837
%     full-network-only 2016-block scenarios (7 sweeps, lite-network dropped);
%     it is no longer a redraw of Table III. Still missing: the n=64 flip-point
%     plot, survival-window timing plot, and PRIM peeling/coverage-density plot.
%  3. STATS: add PRIM box coverage/density and logistic coefficients (see
%     TODO above the RF paragraph).
%  4. RELATED WORK: add 2021-2025 blockchain-governance / fork-empirics /
%     difficulty-adjustment references; anchor the actor taxonomy in
%     peer-reviewed work in addition to BCAP [10].
% ============================================================


\begin{thebibliography}{99}
\small
\bibitem{nakamoto} S.~Nakamoto, ``Bitcoin: A peer-to-peer electronic cash system,'' 2008.
\bibitem{segwit} E.~Lombrozo, J.~Lau, and P.~Wuille, ``BIP 141: Segregated Witness (consensus layer),'' 2015.
\bibitem{taproot} P.~Wuille, J.~Nick, and A.~Towns, ``BIP 341: Taproot: SegWit version 1 spending rules,'' 2020.
\bibitem{warnet} Bitcoin Dev Project, ``Warnet: Bitcoin network testing framework.'' [Online]. Available: \url{https://github.com/bitcoin-dev-project/warnet}
\bibitem{prim} J.~H. Friedman and N.~I. Fisher, ``Bump hunting in high-dimensional data,'' \emph{Statistics and Computing}, vol.~9, no.~2, pp.~123--143, 1999.
\bibitem{bryant} B.~P. Bryant and R.~J. Lempert, ``Thinking inside the box: A participatory, computer-assisted approach to scenario discovery,'' \emph{Technological Forecasting and Social Change}, vol.~77, no.~1, pp.~34--49, 2010.
\bibitem{kristoufek} L.~Kristoufek, ``What are the main drivers of the Bitcoin price? Evidence from wavelet coherence analysis,'' \emph{PLOS ONE}, vol.~10, no.~4, e0123923, 2015.
\bibitem{biais} B.~Biais, C.~Bisi\`ere, M.~Bouvard, and C.~Casamatta, ``The blockchain folk theorem,'' \emph{Review of Financial Studies}, vol.~32, no.~5, pp.~1662--1715, 2019.
\bibitem{hayes} A.~S. Hayes, ``Bitcoin price and its marginal cost of production: support for a fundamental value,'' \emph{Applied Economics Letters}, vol.~26, no.~7, pp.~554--560, 2019.
\bibitem{bcap} R.~Crypto~Fish, S.~Lee, and L.~Alden, ``Analyzing Bitcoin Consensus: Risks in Protocol Upgrades,'' Bitcoin Consensus Analysis Project (BCAP), v1.0, Nov.\ 2024. [Online]. Available: \url{https://github.com/bitcoin-cap/bcap}
\bibitem{gervais} A.~Gervais et al., ``On the security and performance of proof of work blockchains,'' in \emph{Proc. ACM CCS}, 2016, pp.~3--16.
\bibitem{blocksim} M.~Alharby and A.~van Moorsel, ``BlockSim: An extensible simulation tool for blockchain systems,'' \emph{Frontiers in Blockchain}, vol.~3, art.~28, 2020.
\bibitem{simblock} Y.~Aoki, K.~Otsuki, T.~Kaneko, R.~Banno, and K.~Shudo, ``SimBlock: A blockchain network simulator,'' in \emph{IEEE INFOCOM Workshops (CryBlock)}, 2019, pp.~325--329.
\bibitem{neudecker} T.~Neudecker, P.~Andelfinger, and H.~Hartenstein, ``A simulation model for analysis of attacks on the Bitcoin peer-to-peer network,'' in \emph{IFIP/IEEE IM}, 2015, pp.~1327--1332.
\bibitem{decker} C.~Decker and R.~Wattenhofer, ``Information propagation in the Bitcoin network,'' in \emph{IEEE P2P}, 2013.
\bibitem{foytik26} P.~Foytik, ``Scenario potential and surprise-driven scenario discovery in critical-infrastructure settings,'' 2026, in preparation.
\bibitem{eyal} I.~Eyal and E.~G. Sirer, ``Majority is not enough: Bitcoin mining is vulnerable,'' in \emph{Proc. Financial Cryptography (FC)}, 2014.
\bibitem{carlsten} M.~Carlsten, H.~Kalodner, S.~M. Weinberg, and A.~Narayanan, ``On the instability of Bitcoin without the block reward,'' in \emph{Proc. ACM CCS}, 2016, pp.~154--167.
\bibitem{bonneau} J.~Bonneau et al., ``SoK: Research perspectives and challenges for Bitcoin and cryptocurrencies,'' in \emph{IEEE S\&P}, 2015, pp.~104--121.
\bibitem{cong} L.~W. Cong, Z.~He, and J.~Li, ``Decentralized mining in centralized pools,'' \emph{Review of Financial Studies}, vol.~34, no.~3, pp.~1191--1235, 2021.
\bibitem{erlay} G.~Naumenko, G.~Maxwell, P.~Wuille, A.~Fedorova, and I.~Beschastnikh, ``Erlay: Efficient transaction relay for Bitcoin,'' in \emph{Proc. ACM CCS}, 2019, pp.~817--831.
\bibitem{foytik20} P.~Foytik, S.~Shetty, S.~P. Gochhayat, E.~Herath, D.~Tosh, and L.~Njilla, ``A blockchain simulator for evaluating consensus algorithms in diverse networking environments,'' in \emph{Proc. Spring Simulation Conf. (SpringSim)}, 2020, pp.~1--12.
\bibitem{gochhayat} S.~P. Gochhayat, S.~Shetty, R.~Mukkamala, P.~Foytik, G.~A. Kamhoua, and L.~Njilla, ``Measuring decentrality in blockchain based systems,'' \emph{IEEE Access}, vol.~8, pp.~178\,372--178\,390, 2020.
\bibitem{dewolf} L.~de Wolf, \emph{Defending Bitcoin: Industrial-Grade Cybersecurity for the Monetary Grid}, 2026. [Online]. Available: \url{https://defendingbitcoin.com}
\end{thebibliography}
\end{document}